\documentclass[11pt,twoside,a4paper]{article}
\usepackage[left=1.3cm,right=1.3cm,top=1.8cm,bottom=1.8cm]{geometry}
\usepackage{indentfirst}				
\usepackage[english]{babel}			
\usepackage[pdftex]{hyperref}			
\usepackage{lmodern}					
\usepackage{multicol}					
\usepackage{caption}					
\usepackage[font=footnotesize,labelfont=bf]{caption} 
\usepackage{microtype} 					
\usepackage{lipsum}						
\usepackage{subfigure}					
\usepackage{fontawesome}				
\usepackage{xcolor}						
\usepackage{graphicx}					
\hypersetup{colorlinks = true}			
\usepackage{nomencl} 					
\usepackage[utf8]{inputenc}				
\usepackage{lineno}						
\usepackage{float}						
\usepackage{amsmath,array}				
\usepackage{amssymb}					
\usepackage{multirow}					

\begin{document}

\setlength{\parindent}{0.5cm}			
\setlength{\parskip}{0.1cm}  			

\definecolor{blue}{RGB}{41,5,195}		
\makeatletter							
\hypersetup{
    pdftitle={Self-assembly of mixed surfactants sodium dodecylsulfate and polyethylene glycol dodecyl ether in aqueous solutions}, 				
    pdfauthor={Juliano F. Teixeira, Juliana Silva Quintão, Kairon M. Oliveira, Alvaro V. N. C. Teixeira },				
    pdfsubject={Nome da Revista},
    pdfcreator={LaTeX with abnTeX2},
    pdfkeywords={Surfactantes,}{Micelas,}{CMC}, 
    linkcolor=blue,          			
    citecolor=blue,        				
    filecolor=magenta,      			
    urlcolor=blue,
    bookmarksdepth=4
}
\makeatother


\title{Self-assembly of mixed surfactants sodium dodecylsulfate and polyethylene glycol dodecyl ether in aqueous solutions}


\author{Juliano F. Teixeira\thanks{\href{mailto:juliano.f.t@outlook.com}{juliano.f.t@outlook.com}} \and Juliana S. Quintão \and Kairon M. Oliveira  \and Alvaro V. N. C. Teixeira\thanks{\href{mailto:alvaro@ufv.br}{alvaro@ufv.br}}  }

\date{ Departamento de Física, Universidade Federal de Viçosa, Av. Peter Henry Rolfs, sn, Viçosa, 36570-900, Minas Gerais, Brazil }

\maketitle

\vspace{-0.5cm}


\begin{abstract}

The thermodynamic behavior of mixed systems containing the anionic surfactant sodium dodecyl sulfate (SDS) and the nonionic surfactant polyethylene glycol dodecyl ether (Brij L4) in aqueous solutions was investigated. Electrical conductivity and interfacial tension measurements were employed to investigate the concentration-dependent properties of these surfactant mixtures. Two main experiments were conducted: i) constant ratio experiment: the overall surfactant concentration was varied while maintaining a fixed ratio between SDS and Brij L4. It was shown that the critical micelle concentration (CMC) determined from electrical conductivity measurements does not indicate the formation of the first mixed micelles as observed using tensiometry, but the point from which the adsorption of counterions by the existing micelles became important. By using the Regular Solution Theory (RST), it was found that the interaction between SDS and Brij L4 is synergistic, driven by dipole-ion interactions of the hydrophilic regions of the two surfactants. ii) Fixed SDS concentration while increase the Brij L4 concentration: the resulting electrical conductivity exhibited non-trivial behavior which complexity arose from simultaneous phenomena of incorporation of free SDS molecules into the micelles as Brij L4 concentration increased, leading to a decrease in electrical conductivity; liberation of some adsorbed counterions from the mixed micelles to the solution, which increases the conductivity and increase of viscosity due to Brij L4 further addition that leads to a sequential decrease in electrical conductivity.

\noindent \textbf{Keywords}: Surfactants. Micelles. CMC. 

\end{abstract}

\vspace{0.2cm}


\begin{multicols}{2}					

\section{Introduction}
\label{Introduction}

The term “surfactant” (short of “surface active agent”) refers to a group of molecules capable of reducing the surface energy (interfacial tension) of interfaces. Surfactants typically consist of organic molecules with both hydrophilic and hydrophobic parts within the same molecule, which is why they are also classified as amphipathic molecules. While surfactants are widely known as key components in detergents, soaps and cleaning products \cite{YU2008517,SHABAN2020100537}, they are used in a wide range of applications \cite{POSA2023131951, SAH2024124413,PRASADNIRAULA2022120339,kumar2023effect}. For instance, in the pharmaceutical industry, they are utilized to encapsulate and enhance drugs efficiency \cite{islam2023physico,rony2023influences}]. In medicine, surfactants play a crucial role in treating respiratory distress syndrome in newborns \cite{ramanathan2006surfactant} and studies indicates that surfactant contributes to inhibiting multidrug resistance in cancer cells \cite{tang2013key}. Additionally, they are also used in the production of thin films and gold nanoparticles \cite{tatarchuk2017synthesis}. These applications underscore the need for further comprehensive studies on surfactants.

One of the main parameters for characterization of surfactants is the critical micelle concentration (CMC), a concentration in which the surfactant molecules form organized  clusters of molecules in the bulk of the solution, known as micelles.
Surfactant mixtures can be more efficient in certain cases, for example, in petroleum extraction, detergent production \cite{li2016mixtures} and to enhance the solubilization of isoflavonoids \cite{kumar2024binary}. Thus, the study of mixed surfactant systems has been of great interest in the scientific area and technology industry.  Surfactant mixtures have characteristics that are not observed in systems with a single surfactant, like dramatic reduction in CMC, changes in viscosity and others \cite{rubingh1979solution}. These phenomena can be investigated with measurements of interfacial tension, viscosity, electrical conductivity, among others. 

\begin{figure}[H]
\centering
\includegraphics[width=0.95\linewidth]{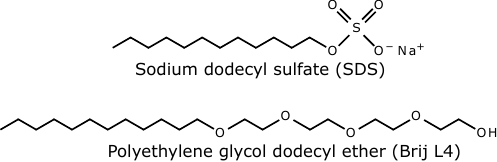}
\caption{ Estrutura química do SDS e Brij L4. }
\label{Estrutura}
\end{figure}

One of the most extensively studied surfactant is the anionic surfactant sodium dodecyl sulfate (SDS) and its interaction with other ionic and nonionic surfactants \cite{zhang2005interaction,mazen2008mixing,ABBOT2021116352}. However, to the best of our knowledge, there has been no study evaluating the individual effect of SDS and a second surfactant on the mixed micelle formation.
We investigated the interactions between the anionic surfactant SDS and the nonionic surfactant polyethylene glycol dodecyl ether (Brij L4). Both surfactants have the same hydrophobic chain with twelve carbons atoms. SDS has a sulfate group as hydrophilic part and Brij L4 a polyethylene glycol chain with four ethylene glycol units (\autoref{Estrutura}). 
A systematic study of mixed SDS and Brij surfactants in aqueous solution was realized in three main experiments: first, the total surfactant concentration was varied while the ratio of the surfactants was kept constant. On the second study, SDS concentration was varied while the Brij concentration was kept constant. Lastly, the SDS concentration was kept constant and Brij concentration was varied. 
   
\section{Experimental Section}
\label{Experimental Sections}

\subsection{Materials}
\label{Materials}

Sodium dodecyl sulfate, SDS (\textit{M} = 288.3 g/mol, $99\,\%$ purity, CAS number 436143, MDL number MKCD4141 and MKCD7976) and polyethylene glycol ether, Brij L4  (\textit{M} = 362 g/mol, CAS number 235989, MDL number, MKBT1319V)  manufactured by Sigma-Aldrich, St. Louis, USA. All chemicals were used without further purification.

Surfactant solutions were prepared using deionized water (resistivity of $18.2$ M$\Omega $.cm). Stock solutions were homogenized and left overnight to ensure the equilibrium  was achieved.

\subsection{Electrical conductivity measurements}
\label{Electrical conductivity measurements}

The electrical conductivity of the solutions was measured using a digital conductivity meter, model DM32, manufactured by Digimed, São Paulo, Brazil. The measurements were conducted with a dip-type conductivity cell, model DMC-010M, made of a platinum disk with a cell constant K = 1 cm$^{-1}$. Titrations of stock solutions into the main solution were made by Harvard Apparatus PHD syringe pump in order to vary the surfactant concentration.  Typically, 125 injections  with 40 $\mu$L each one were made to achieve the desirable concentration. After each injection, the system was left under agitation for at least 30 seconds using a magnetic bar and a magnetic stirrer. After this time, electrical conductivity of the solution  were measured  every 1 second and the measurement was considered completed when 10 consecutive measurements had the same value within the accuracy of the device. At this point, we considered that the system has reached equilibrium.  Thus, we obtained the profile of electrical conductivity with the concentration of surfactant. All experiments were carried out in a thermal bath keeping the temperature controlled at ($25.0 \pm 0.1$) $^{\circ}$C.  Number and volume of injections, speed of agitation and agitation time were controlled by a homemade program in \textit{LabView}. The communication between the software and all the hardware: syringe pump, conductivity meter and magnetic stirrer were made by  using an Arduino board.

\subsection{Interfacial tension measurements}
\label{Interfacial tension measurements}

The interfacial tension was measured using a Tensiometer,  model DCTA 21, manufactured by \textit{Dataphysics}, Filderstadt, Germany.  The measurement was conducted applying the Wilhelmy plate method with a value of 0.05 mN/m as the standard deviation limit.  The operation of the tensiometer is based on the force required to separate a platinum-iridium  plate in contact with the surface solution.
The interfacial tension is linked to the CMC of the surfactant, since micelles are formed in the bulk of solution after the surface  is saturated.
The measurements were carried out at a room temperature of $(23 \pm 1)$ $^{\circ}$C. After each addition of surfactant the solution was gently homogenized to avoid foaming.

\subsection{Viscosity measurements}
\label{Viscosity measurements}

A Cannon-Fenske routine viscometer was used to determine the dynamic viscosity of the solutions.  The viscometer was kept in a thermal bath in order to control the temperature at ($25.0 \pm 0.1$) $^{\circ}$C. The dynamic viscosity ($\eta$) was calculated from the measurement of the time ($\Delta t$) required for fluid flow between two chambers connected by a capillary with 0.63 mm thickness  and the density ($\rho$) of the solution as shown in  \autoref {calcvisc}:

\begin{equation}
\eta = \rho \delta( \Delta t - \upsilon),
\label{calcvisc}
\end{equation}

\noindent where  $ \delta = 0.0150 \, \textrm{m}^2 / \textrm{s}^2 $ and $\upsilon = 2732\,\textrm{s}^3 / \Delta t ^2$ is a time correction factor. 
The density of the solution was measured by an oscillatory densiometer (Schmidt Haensch, EDM, Germany) at a temperature of ($25.00 \pm 0.05$) $ ^{\circ}$C.

\section{ Results and Discussion}
\label{Results and Discussion}

A systematic study of mixed SDS and Brij surfactants in aqueous solution was realized in three main experiments: first, varying the total surfactant concentration ([S] = [SDS] + [Brij]) keeping constant the ratio of the surfactants ($\alpha = \textrm{[SDS]}/\textrm{[S]}$) (\autoref{Constant ratio of surfactants}). On the second study [SDS] was varied keeping [Brij] constant (\autoref{Constant Brij concentration}). Lastly, [Brij] was varied keeping [SDS] constant (\autoref{Constant SDS concentration}). Here square brackets are used to indicate the molarity of a substance.

\subsection{Constant ratio of surfactants} 
\label{Constant ratio of surfactants}

\begin{figure*}[htb]
\centering
\includegraphics[width=\linewidth]{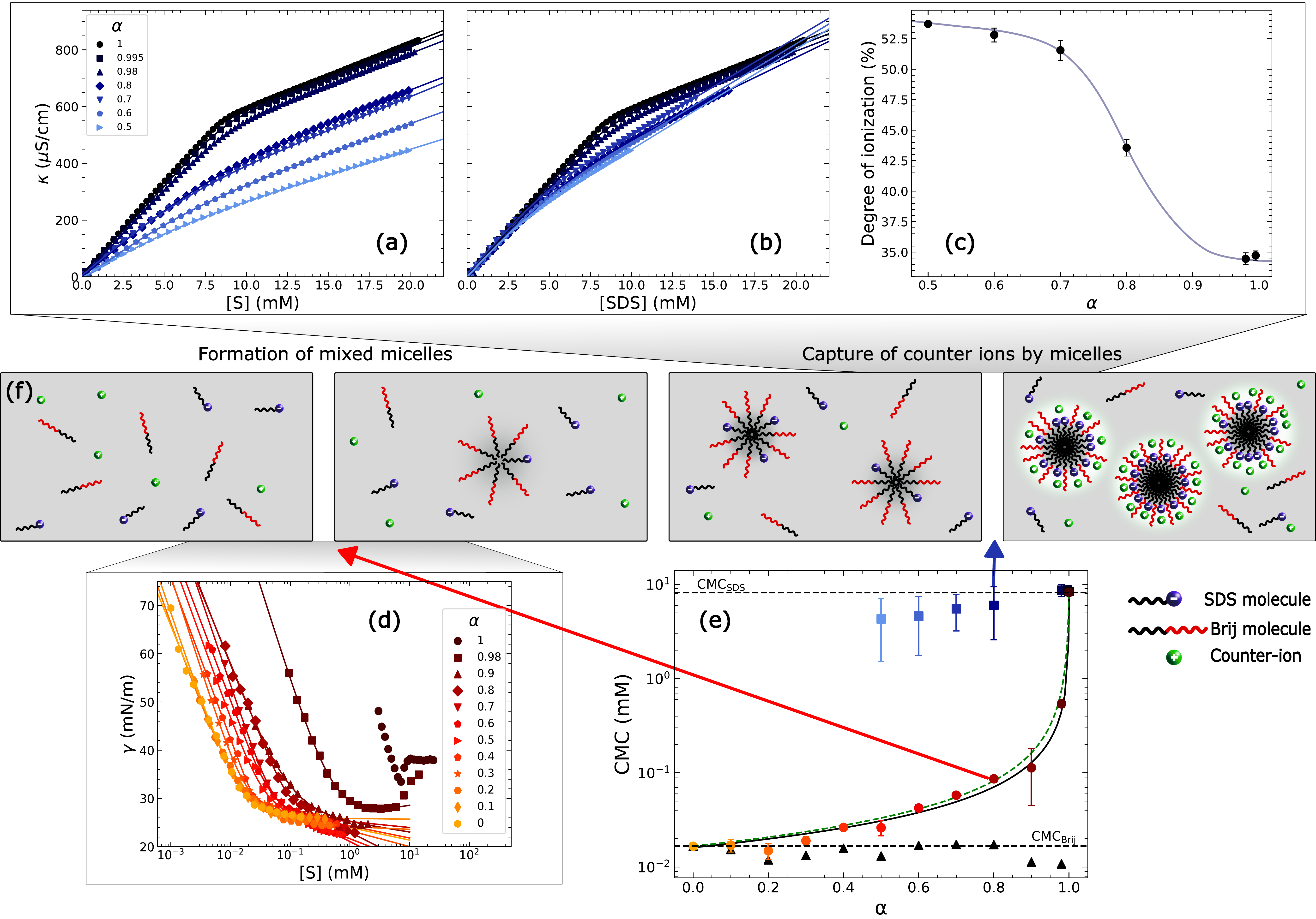}
\caption{  Electrical conductivity versus \textbf{(a)} total surfactant concentration and \textbf{(b)} [SDS] for different values of $\alpha$. Continuous lines are the fit of the experimental data by \autoref{FIT}. \textbf{(c)} Degree of ionization of micelles at CMC for different values of $\alpha$. The continuous line is a guide to the eye. \textbf{(d)} Interfacial tension versus total surfactant  concentration for different values of $\alpha$. Continuous lines are the fit of the experimental data by \autoref{eqTI} \textbf{(e}) CMC of a mixed surfactant solution with SDS and Brij versus $\alpha$ from: electrical conductivity measurements (\scriptsize \faStop) and surface tension measurements (\scriptsize \faCircle); Brij concentration at tensiometric CMC ($ \blacktriangle$). Dashed line represents ideal mixing given by Regular Solution Theory (\autoref{RSTideal}).   Continuous line represents the theoretical behavior  for $\beta = -3.2$. Dashed horizontal lines represents the $\textrm{CMC}_{\textrm{SDS}}$ and $\textrm{CMC}_{\textrm{Brij}}$. \textbf{(f)} Schematic illustration of mixed micelle formation process and capture of counterions.  }
\label{Fig2}
\end{figure*}

The electrical conductivity ($\kappa$) for the mixture SDS and Brij at different $\alpha$ values (fraction of SDS) is presented in  \autoref{Fig2}a. Electrical conductometry is a widely used method to find the CMC of ionic surfactants, characterized by the point where the slope of the linear growing of the electrical conductivity changes. The well-known behavior of the two linear regimes is due mainly to the mobility of the counterions that increases linearly with the surfactant concentration. If the concentration is high enough to form micelles, part of the counterions are bound to the micelles, decreasing the slope. The ratio of the slopes gives the degree of ionization \cite{pashley2004applied}.

The same behavior was observed for systems with low amount of Brij ($0.8 < \alpha \le 1$). For $\alpha = 1$ (only SDS) it was found $\textrm{CMC}_{\textrm{SDS}} = 8.4 \pm 0.2$ mM and a degree of ionization of 35 \%  which agrees with the values reported in literature \cite{PAL201342}. With the addition of further Brij to the mixture ($\alpha \le 0.8$) the transition and the slopes in the pre and post micellar regime becomes less evident. This means that the onset of micelle formation does not occur in a narrow range of concentration, but it appears gradually as the surfactant concentration increases. This effect was also observed in other systems with SDS when adding a cosolvent to the solution and the gradual transition was also interpreted as gradual formation of micelles throughout the concentration range \cite{vsteflova2016comprehensive}. On the other hand the physical explanation here is different: since the electrical conductivity is due mainly to the mobility of the counterions it is reasonable to assume that the increase in the total surfactant concentration leads to a continuous incorporation SDS molecules to the micelles, and so some of the counterions. This should occur even at low concentrations since critical the micellar concentration of Brij is small ($\textrm{CMC}_{\textrm{Brij}} \sim 2 \times 10^{-2}$ mM). Representing the same data in terms of [SDS] only (=$\alpha \times [S]$) and it can be seen that the initial slope is the same for all systems. This shows that the initial electrical conductivity growth is due to the mobility of the free counterions only and it is not influenced by Brij (\autoref{Fig2}b).  These results also indicate that the contribution of SDS molecules to the total conductivity is negligible since the conductivity of the solution at $\alpha = 1$ (with SDS molecules in free form) is the same as for lower $\alpha$, where some SDS is incorporated into the mixed micelles.

CMC values were determined by adjusting the experimental data following the methodology proposed by the reference \cite{carpena2002problems} where the derivative of the electrical conductivity  ($d\kappa/d\textrm{[S]}$) is fitted by a sigmoid function. Alternatively $\kappa$ can be fitted by the integral of sigmoid function (\autoref{FIT}):

\begin{eqnarray}
\kappa (\textrm{[S]}) &=&  \kappa (0) + A_1 \textrm{[S]} + \Delta (A_2 - A_1) \times \nonumber \\
&& \ln\left[\frac{1+\exp\left(\frac{\textrm{[S]}-\textrm{CMC}}{\Delta}\right)}{1+\exp\left(\frac{-\textrm{CMC}}{\Delta}\right)}\right] \textrm{,}
\label{FIT}
\end{eqnarray}

\noindent where $\kappa (0)$ is the electrical conductivity at [S] = 0 mM, $A_1$ and $A_2$ are the slopes before and after CMC and $\Delta$ is the concentration range of the transition between the two linear behaviors.
For small $\alpha$ the transition is not well-defined and the values found for CMC are not reliable. Still, the apparent CMC decreases with increasing [Brij] (\autoref{Fig2}e). The errors bars were estimated as $\pm \Delta$/2. The degree of ionization was estimated by $A_{2}/A_{1}$ fixing the initial slope for all curves for the one obtained by $\alpha=1$ and also by linear fit on initial and final points in each curve. The error bar was estimated by the deviation of the two values. It can be seen in (\autoref{Fig2}c) that micelles in mixtures richer in SDS (larger $\alpha$) has a lower degree of ionization, which can be understood since larger $\alpha$ means that the density of SDS increases and so the electrostatic attraction between micelles and counterions. An important reduction in this parameter is observed at $\alpha \sim 0.8$ which indicates that the association of SDS in the micelles is stronger from this point. The same result was found analyzing the composition of the micelles and will be shown latter. Due to hydrogen bonding and excluded volume the degree of ionization of Brij micelles are not 100 \% \cite{yekymov2022effects}.

Tensiometric technique was also used to determine the CMC.
\autoref{Fig2}d shows the interfacial tension of the mixtures varying the total surfactant concentration for different values of $\alpha$. A similar two-regimes separated by the formation of micelles is observed using tensiometry and it is explained by the continuous accommodation of surfactant molecules on the surface  until the formation of micelles in the bulk which change the accommodation of further molecules on the surface and so the surface tension \cite{sahu2015anionic}.
For $\alpha = 1$ and $\alpha = 0.98$ it can be seen that the interfacial tension increases its value after reaching a  minimum  which is associated to the presence of impurities in the SDS \cite{shedlovsky1949effect}.

The CMC were determined by  \autoref{eqTI}, which is analogous to  \autoref{FIT} but using the logarithmic of the concentration.

\begin{eqnarray}
\gamma(\textrm{[S]})
&=& \gamma_{0} + A_1 \textrm{log([S]}) + \Delta (A_2 - A_1) \times \nonumber \\
&& \ln\left\lbrace \frac{1+\exp\left[\frac{\textrm{log}(\textrm{[S]}/\textrm{CMC})}{\Delta}\right]}{1+\exp\left[\frac{-\textrm{log}(\textrm{CMC})}{\Delta }\right]}\right\rbrace .
\label{eqTI}
\end{eqnarray}

\noindent CMC values obtained by interfacial tension and electrical conductivity are shown in \autoref{Fig2}e. The error bars were estimated either by the error in fitting the data  or by the standard deviation for repetitions of the same measurement.

\autoref{Fig2}e indicates that CMC values obtained by electrical  conductivity are one to two orders of magnitude higher than the ones by tensiometry. This difference comes from the fact that the conductivity is sensitive to the counter-ion mobility while tensiometry is sensitive to the coating of the air-water interface by surfactant molecules. Thus, tensiometry give us the real CMC while the conductometry gives the point from which the already existed micelles starts to behave more like an ionic micelle and adsorbs more effectively the free counterions.

\subsubsection{Regular Solution Theory} 
\label{Regular Solution Theory}

We used the Regular Solution Theory (RST) proposed by Rubingh to predict the behavior of micellization in binary surfactant solution with  nonionic surfactant (NI) and anionic surfactant (A) \cite{rubingh1979solution, sahu2015anionic}. According to RST, the critical micellar concentration of a mixed solution  $ \textrm{CMC}  $ is given by  \autoref{RST}:
 
\vspace{1cm}
\begin{equation}
\frac{1}{\textrm{CMC} } = \frac{\alpha}{f_{\mbox{\scriptsize A}} \mbox{CMC}_{\mbox{\scriptsize A}}} + \frac{1-\alpha}{f_{\mbox{\scriptsize NI}}\mbox{CMC}_{\mbox{\scriptsize NI}}} ,
\label{RST}
\end{equation}

\noindent where $\alpha$ is the mole fraction of the anionic surfactant which is expressed as $\alpha = \textrm{[A]}/ \textrm{[S]}$, where [A] is the anionic surfactant concentration and [S] is the total surfactant concentration. The parameters $f_{\mbox{\scriptsize A}}$ and $f_{\mbox{\scriptsize NI}}$ are the activity coefficients of the anionic and the nonionic surfactants, respectively, and their values indicate the deviation from ideality in mixtures. $\mbox{CMC}_{\mbox{\scriptsize A}}$ and $\mbox{CMC}_{\mbox{\scriptsize NI}}$ are the critical micellar concentration of the anionic and the nonionic surfactant. In an ideal mixture $f_{\mbox{\scriptsize A}} = f_{\mbox{\scriptsize NI}} = 1$ and \autoref{RST} reduces to  \autoref{RSTideal} as follows:

\begin{small}
\begin{equation}
\frac{1}{\textrm{CMC} } = \frac{\alpha}{\mbox{CMC}_{\mbox{\scriptsize A}}} + \frac{1-\alpha}{\mbox{CMC}_{\mbox{\scriptsize NI}}}.
\label{RSTideal}
\end{equation}
\end{small}

\noindent The type and extent of interaction between surfactants can be quantitatively expressed by the sign and magnitude of the interaction parameter $\beta$ (\autoref{b}). A {\it synergism}, a term used to describe a thermodynamically favorable interaction between different types of surfactants, happens when $\beta$ is negative. 
For a positive $\beta$, an {\it antagonism} takes place, indicating that the more favorable interaction is between molecules of the same surfactant. 
Lastly, when $\beta$ is zero, the mixing is ideal, indicating that there is no distinction in the affinity between different types of surfactants.

From the RST it is possible to relate the molar fraction of anionic surfactant ($X_{\mbox{\scriptsize A}}$\footnote[2]{Both $\alpha$ and $X_{\mbox{\scriptsize A}}$ correspond to the molar fraction of anionic surfactant in relation to the total amount of surfactants. However, $\alpha$ is the molar fraction of SDS in the solution and $X_{\mbox{\scriptsize A}}$ is the molar fraction of SDS inside the micelle.}) with the other parameters as shown in \autoref{XA}.

\begin{equation}
X_{\mbox{\scriptsize A}}^2 \ln \left(\frac{ \alpha \textrm{CMC}}{X_{\mbox{\scriptsize A}} \mbox{CMC}_{\mbox{\scriptsize A}} }\right) = X_{\mbox{\scriptsize NI}}^2 \ln \left[\frac{(1 - \alpha) \textrm{CMC} }{ X_{\mbox{\scriptsize NI}} \mbox{CMC}_{\mbox{\scriptsize NI}}} \right],
\label{XA}
\end{equation}

\noindent with $X_{\mbox{\scriptsize NI}}= 1 - X_{\mbox{\scriptsize A}}$. After obtaining $X_{\mbox{\scriptsize A}}$ numerically it is possible to calculate  $\beta$, $f_{\mbox{\scriptsize A}}$ and $f_{\mbox{\scriptsize NI}}$ from \autoref{b}, \ref{fa} and \ref{fni}.

\begin{equation}
\beta = \frac{1}{\left( 1- X_{\mbox{\scriptsize A}}\right)^2} \ln \left(  \frac{\alpha \textrm{CMC}}{X_{\mbox{\scriptsize A}} \mbox{CMC}_{\mbox{\scriptsize A}} } \right).
\label{b}
\end{equation}

\begin{equation}
f_{\mbox{\scriptsize A}} = \exp \left[ \beta \left(1-X_{\mbox{\scriptsize A}}\right)^2 \right],
\label{fa}
\end{equation}

\begin{equation}
f_{\mbox{\scriptsize NI}} = \exp \left[ \beta (X_{\mbox{\scriptsize A}})^2 \right].
\label{fni}
\end{equation}

The activity coefficients are related to the deviation from ideality for each one of the surfactants, as can be seen in the  \autoref{RST} and \autoref{fa} and \ref{fni} show us that the richer the micelle from one of the surfactants, the closer to one is the activity coefficient related to this surfactant. The parameter $\beta$ appears as a magnifier of this tendency.

\begin{figure*}[htb]
\centering
\includegraphics[width=0.85\linewidth]{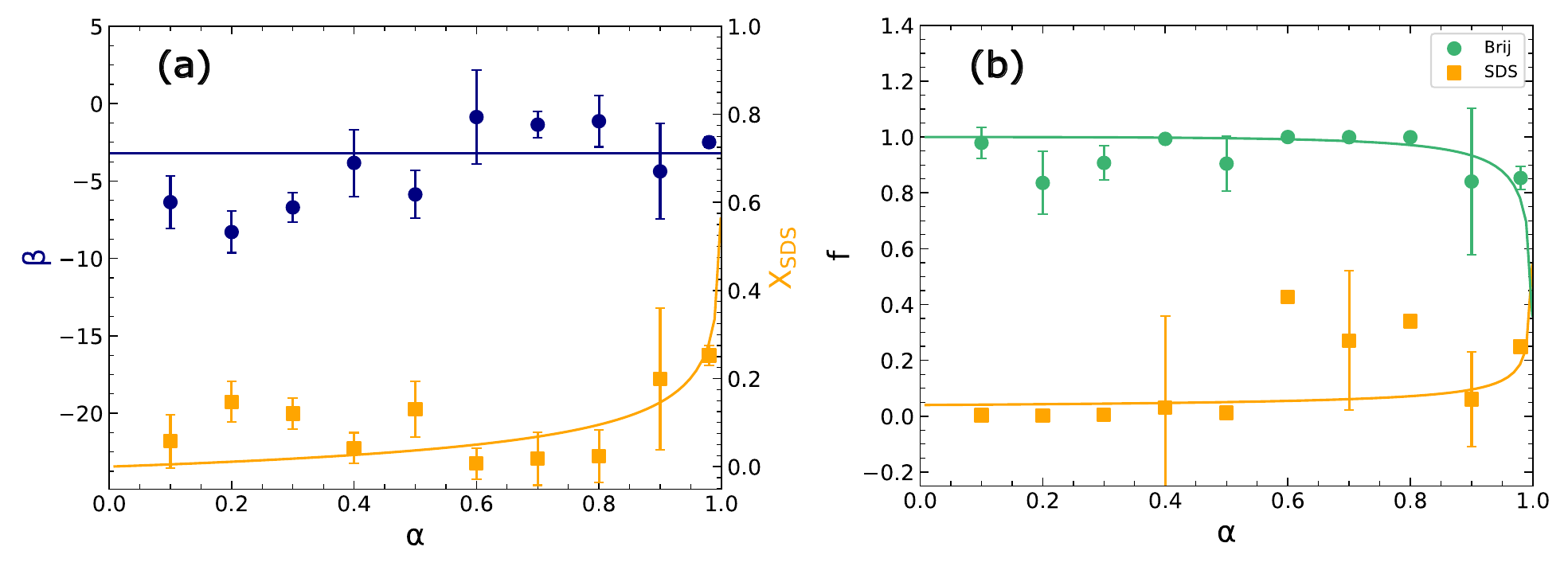}
\caption{\textbf{(a)} Interaction parameter $\beta$ and SDS molar fraction in the micelle    for different $\alpha$. The continuous lines represent the theoretical behavior for constant $\beta = -3.2$. \textbf{(b)} SDS and Brij activity coefficients for different $\alpha$ values. Continuous line represent the theoretical behavior for constant $\beta$ equal to $-3.2$.
}
\label{Fig3}
\end{figure*}

The parameters calculated from the tensiometry are shown in  \autoref{Fig3}a and \autoref{Tab1}.

\autoref{Fig3}a suggests that $\beta$ has a weak dependence with $\alpha$, but this statement is unreal since $\beta$ is a parameter that gives a quantitative value of the synergism/antagonism in the mixed micelle formation and it is independent of temperature, concentration and surfactants ratio \cite{abe2004mixed}.
Taking the average value for all concentrations, we found $\beta = -4 \pm 1$. On the other hand, comparing the predicted curves of CMC, $X_\textrm{SDS}$, $f_{\textrm{SDS}}$ and $f_{\textrm{Brij}}$ (plotted as continuous lines on Figures \autoref{Fig2}e, \ref{Fig3}a and \ref{Fig3}b) with the experimental results we found that the best value was $\beta=-3.2$. The effect of different values of $\beta$ does not impact on the theoretical CMC curve as it does in $X_\textrm{SDS}$ and $f$. \autoref{Fig7} shows the theoretical values of these parameters for $\beta$ ranging from 0 to -5, along with the experimental data. The value of $\beta$ agrees with the one for mixed system of cationic and nonionic surfactants ($ - 5 \leq  \beta \leq -1$). For anionic-nonionic surfactants even more negative values are found. The stronger interaction is for anionic and cationic surfactants ($\beta \leq -10$) \cite{holland1992mixed}.

The negative value of $\beta$ implies that these surfactants exhibit a synergistic interaction, meaning that the SDS-Brij interaction is thermodynamically more favorable than Brij-Brij and SDS-SDS. This synergy is explained by the fact that the hydrophilic chain of the nonionic surfactant involves the charged head of the anionic surfactant partially shielding the electrostatic repulsion and favoring micelle formation, which is what it happens between SDS and Brij surfactants \cite{sahu2015anionic}.

Activity coefficients for SDS ($f_{\textrm{SDS}}$) and Brij ($f_{\textrm{Brij}}$) with $\alpha$ are shown in  \autoref{Fig3}b. We can observe that $f_{\textrm{Brij}} \approx 1$ and $f_{\textrm{SDS}} \approx 0$ for $\alpha \lesssim 0.8$. This shows that the mixed micelle in this range are practically formed by Brij molecules. The same result is seen by calculating  $X_{\textrm{SDS}}$ as it is close to zero for $\alpha \lesssim 0.8$ (\autoref{Fig3}a). Finally, the concentration of Brij in the onset of formation of micelles (CMC) is approximately equals to the $\textrm{CMC}_{\textrm{Brij}}$ ($= 1.5 \times 10^{-2}$ mM) for $\alpha \lesssim 0.8$ confirming again that the micelles are formed primarily of Brij molecules (\autoref{Fig2}e and \autoref{Fig3}b).

For $\alpha > 0.8$ we can observe that [Brij] decreases at the CMC showing that SDS starts to have a significant participation on the micelles' composition. Similarly, for $\alpha > 0.8$ it was observed an increase of $X_{\textrm{SDS}}$  (\autoref{Fig3}a) and deviations of $f_{\textrm{Brij}}$ and $f_{\textrm{SDS}}$ (\autoref{Fig3}b). Finally, the importance of SDS in the micelles and so the increase of is surface charge for $\alpha > 0.8$ is reflected at the drop in the degree of ionization in  \autoref{Fig2}c.

\subsubsection{Thermodynamic characterization} 
\label{Thermodynamic characterization}

Several thermodynamic parameters can be evaluated from the data extracted by the conductometric and tensiometric techniques and their mean values with $\alpha$ are presented in \autoref{Tab2}. Most of the results are not precise enough to make an accurate analysis of their dependence with $\alpha$ (see Supplementary Information for the overall results).

It was found that the surfactant concentration adsorbed in air/water interface at CMC ($\Gamma_{\textrm{CMC}}$) seem to increase with $\alpha$ (which is equivalent to the decrease of the average area occupied by a single molecule at the interface, $A_{\textrm{avg}}$). Since the hydrophobic chain of SDS and Brij are the same, these results are due to the differences of the hydrophilic part. There are two factors that explain this behavior: the bigger volume and so the steric repulsion of Brij hydrophilic part; and additional electric repulsion due to polarization along the chain \cite{duan2011super,duan2012super}, making the  $\Gamma_{\textrm{CMC}}$ of  Brij smaller and $A_{\textrm{avg}}$ bigger than SDS.

\begin{table*}[htb]
\centering
\caption{Surface pressure ($\Pi_{\textrm{CMC}}$), surface excess concentration ($\Gamma_{\textrm{CMC}}$), average area occupied by a single molecule at the interface ($A_{\textrm{avg}}$), free energy at the air/interface ($G_{\mbox{{\scriptsize min}}}$), variation on free energy during the micellization ($\Delta G_{m}$), variation of free energy change of adsorption ($\Delta G_{\textrm{ads}}$) for different $\alpha$. }

   \begin{tabular}{>{\centering\arraybackslash}p{0.7cm}  >{\centering\arraybackslash}p{1.2cm}  >{\centering\arraybackslash}p{1.8cm}  >{\centering\arraybackslash}p{1.6cm}  >{\centering\arraybackslash}p{1.4cm}  >{\centering\arraybackslash}p{2.2cm} 
>{\centering\arraybackslash}p{1.6cm} }
  \hline
\textbf{$\alpha$} &	\textbf{$\Pi_{\textrm{CMC}}$ (nN/m)}&  \textbf{$\Gamma_{\textrm{CMC}}$ $\times 10^{-7}$ (mol/$\textrm{m}^2$)} & \textbf{$A_{\textrm{avg}} \, \times 10^{-20} \, (\textrm{m}^2)$ } & \textbf{$G_{\mbox{{\scriptsize min}}}$ kJ/mol} & \textbf{$\Delta G_{m}$ kJ/mol} & \textbf{$\Delta G_{\textrm{ads}}$ kJ/mol}	\\  \hline

0		&	$42 \pm 3$	&	$10 \pm 1$	 &	$172 \pm 11$				&	$32 \pm 5$				&		$-27.3 \pm 0.1$ & 	$-70 \pm 5$			\\ [2pt]

0.1		&	$40 \pm 4$	&	$11 \pm 1$	 &	$155 \pm 13$	&	$31 \pm 6$	&	$-27.2 \pm 0.02$  & $-64 \pm 7$		\\ [2pt]

0.2		&	$39 \pm 4$	&	$11 \pm 1$	&	$156 \pm 14$	&	$31 \pm 6$	&	$-27.6 \pm 0.3$ & $-64 \pm 7$	\\ [2pt]

0.3		&	$39 \pm 3$	&	$13 \pm 1$	&	$131 \pm 9$	&	$26 \pm 4$	&	$-27.0 \pm 0.2$	& $-57 \pm 5$ \\ 

0.4		&	$43 \pm 2$	&	$10.5 \pm 0.4$	&	$159 \pm 7$		&	$29 \pm 3$		&	$-26.1 \pm 0.1$	& $-67 \pm 4$\\ 

0.5		&	$40 \pm 3$	&	$14 \pm 2$ &	$121 \pm 12$		&	$24 \pm 5$		&	$-26.2 \pm 0.3$ & $-55 \pm 5$	\\ 

0.6		&	$41 \pm 3$	&	$11 \pm 1$	&	$145 \pm 8$		&	$28 \pm 4$			&	$-25.0 \pm 0.2$ & $-61 \pm 4$	\\ 

0.7		&	$41 \pm 4$	&	$11 \pm 1$	&	$150 \pm 10$	&	$28 \pm 5$		&	$-24.2\pm 0.1$ & $-61 \pm 6$	\\ 

0.8		&	$41 \pm 3$	&	$11 \pm 1$	&	$156 \pm 9$		&	$29 \pm 4$			&	$-23.2 \pm 0.2$	& $-62 \pm 5$\\ 

0.9		&	$39 \pm 10$		&	$11 \pm 4$	&	$149 \pm 46$		&	$30 \pm 18$		&	$-23 \pm 1$ & $-57 \pm 20$	\\ 

0.98	&	$34 \pm 1$		&	$11 \pm 1$	&	$154 \pm 18$	&	$31 \pm 5$		&	$-18.7 \pm 0.2$ & $-55 \pm 5$	\\ 

1		&	$35 \pm 0$	&	$14 \pm 0$	 &	$119 \pm 0$				&	$27 \pm 0$					&		$-11.8 \pm 0$	& $-37 \pm 0$			\\ 
\hline

\end{tabular}
\label{Tab2}
\end{table*}

Additional characterization was made by means of the free energy at the interface at the air-water interface ($G_{\mbox{{\scriptsize min}}}$), the variation of free energy of micellization ($\Delta G_{m}$) and the variation of free energy of  adsorption  ($\Delta G_{\textrm{ads}}$).

\subsection{Constant Brij concentration} 
\label{Constant Brij concentration}

In this second part electrical conductivity measurements were realized on solutions with  Brij concentration kept constant and varying the SDS concentration.

\begin{figure*}[htb]
\centering
\includegraphics[width=0.85\linewidth]{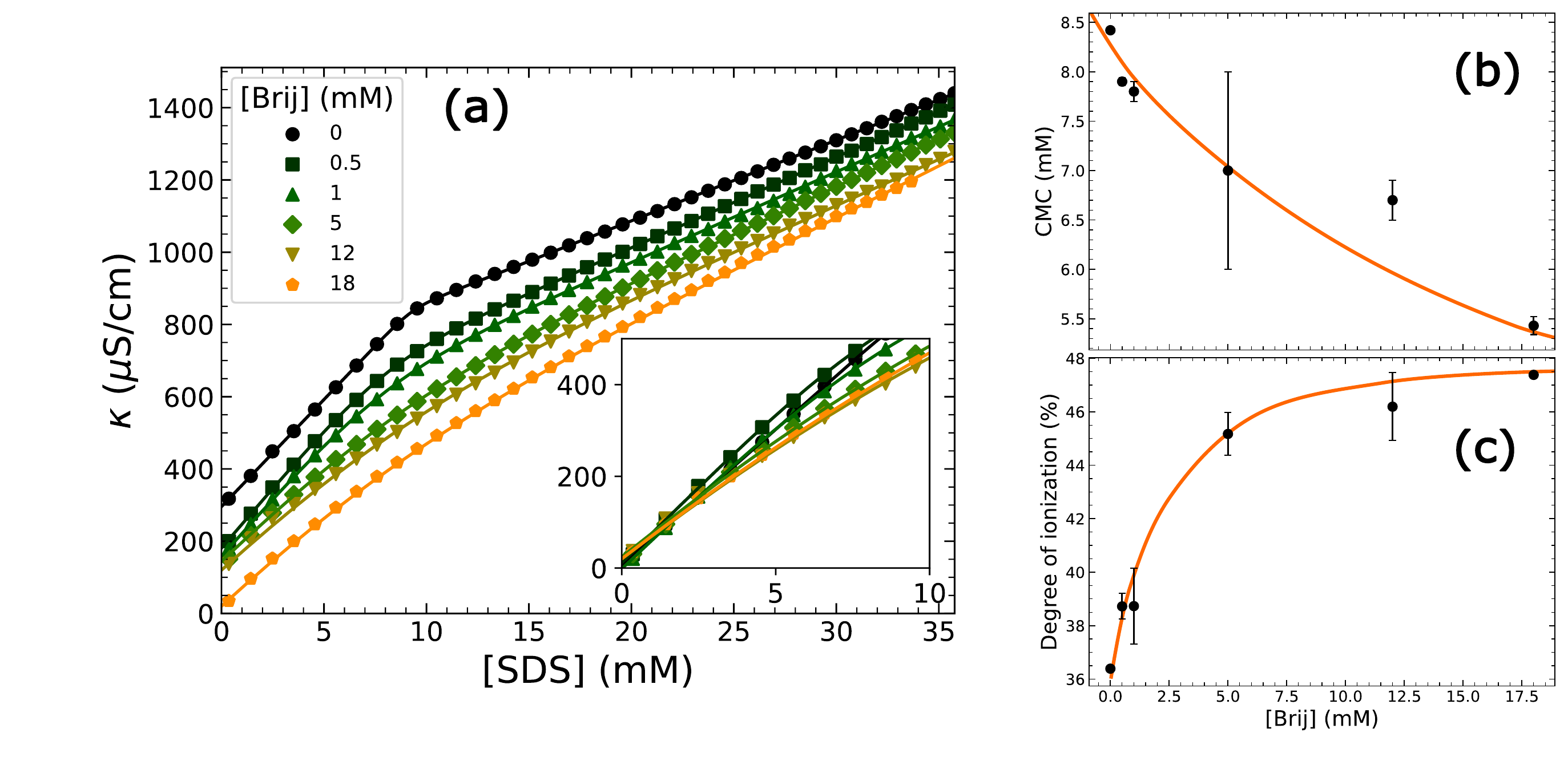}
\caption{\textbf{(a)} Electrical conductivity versus [SDS] for different solutions with [Brij] kept constant. The curves have been vertically shifted for better visualization. Detail: the same results without the vertical shift, showing that the slope for low concentrations is the same. \textbf{(b)} CMC of a mixed surfactant solution with SDS and Brij versus [Brij]. The continuous line is a guide to the eye. \textbf{(c)} Degree of ionization of micelles at CMC for different [Brij] kept constant.
}
\label{Fig4}
\end{figure*}

\autoref{Fig4}a shows the electrical conductivity increasing with SDS concentration, keeping the [Brij] constant. The behavior is similar as to one observed before: two linear regimes with the transition being damped as more Brij is added. It can be observed that the initial slope is the same for all [Brij], showing that the presence of Brij do not affect the mobility of the counterions and so the conductivity for smaller [SDS], even with the presence of mixed micelles, indicating that there are no attached counterions in the micelles for small amounts of SDS.  

The conductometric CMC was determined as before and it gives the [SDS] from which the charge of the micelles are big enough to re-adsorb part of the counterions. \autoref{Fig4}b and  \autoref{Fig4}c show that the conductometric CMC decreases and the degree of ionization increases with increasing of [Brij]. This is a consequence of the synergism between the surfactants that promotes greater transfer of more free SDS molecules into the micellar form and also the dilution of electrical charge of the micelles surface as the [Brij] increases.

\subsection{Constant SDS concentration}
\label{Constant SDS concentration}

\begin{figure*}[htb]
\centering
\includegraphics[width=\linewidth]{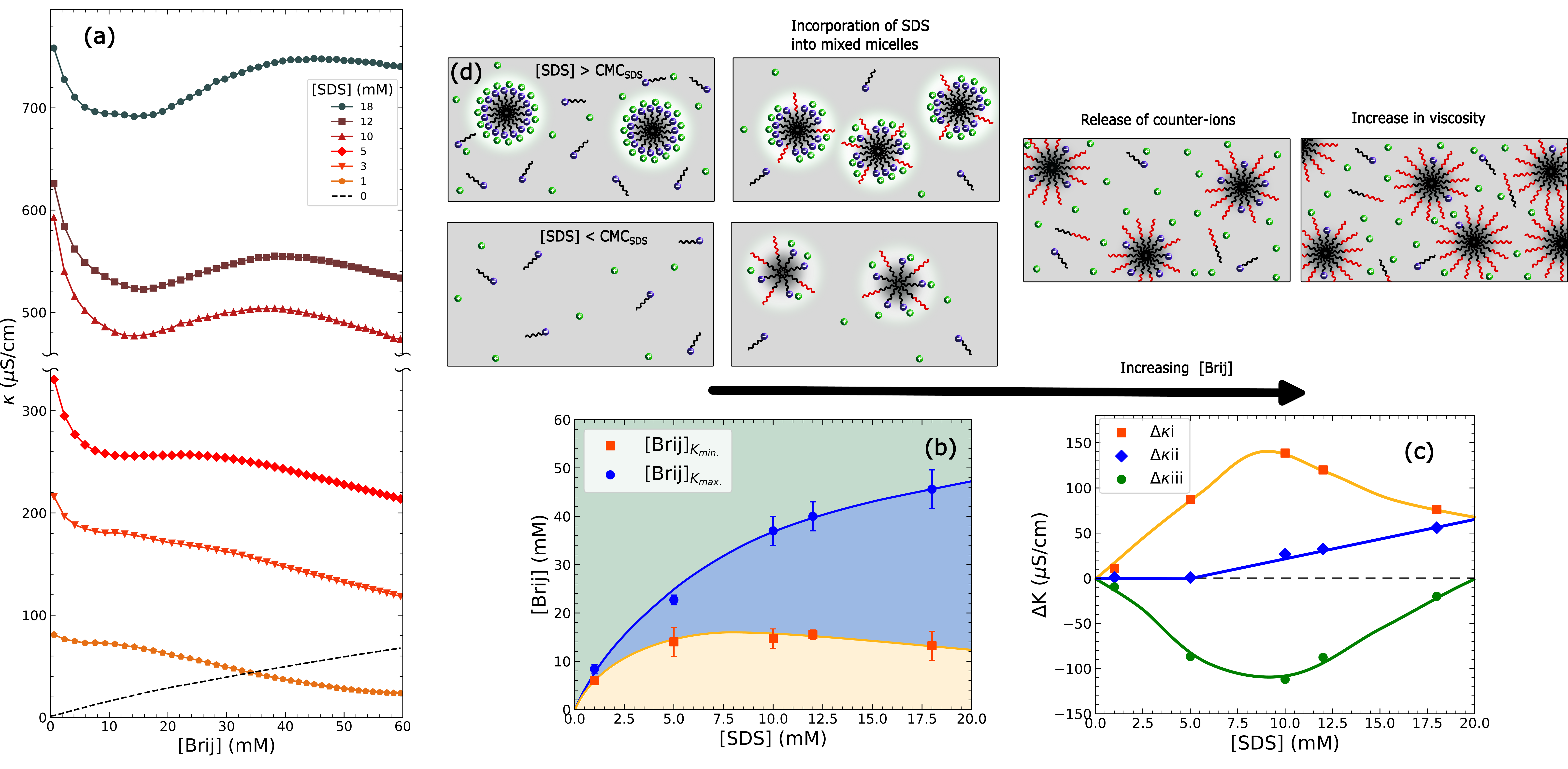}
\caption{\textbf{(a)} Electrical conductivity of aqueous solution varying [Brij]. Corrected electrical conductivity for different solutions with [SDS] kept constant. \textbf{(b)}  [Brij] corresponding to maximum and minimum electrical conductivity for different solutions with [SDS] kept constant. The continuous line is a guide to the eye. \textbf{(c)}  Range from the minimum to initial electrical conductivity ($\Delta \kappa\textbf{i}$); range from the minimum to maximum electrical conductivity ($\Delta \kappa\textbf{ii}$) and range from the initial to maximum electrical conductivity ($\Delta \kappa\textbf{iii}$) for different solutions with [SDS] kept constant. The continuous line is a guide to the eye. \textbf{(c)} Schematic illustration of capture of counterions mixed micelle formation, follow by dilution of SDS in micelle and release of counterions and the increase in the viscosity.
}
\label{Fig5}
\end{figure*}

In this section, we show the results of electrical conductivity and dynamic viscosity measurement for solutions changing [Brij] from 0 to 60 mM while [SDS] was kept constant. The results are shown in \autoref{Fig5}a for [SDS] = 0, 1, 3, 5, 10, 12 and 18 mM. For the curve with [SDS] = 0 mM (dashed line in  \autoref{Fig5}a) it can be seen that the conductivity increases adding Brij. This effect can be explained due to the ionization of water molecules from the polyethylene glycol chains and subsequential formation of a pseudo-polycation by association of H$^{+}$ \cite{duan2011super, duan2012super}. 
From the results of reference \cite{duan2011super} it is shown an increase in conductivity of approximately 4 $\mu$S/cm per mol of ethylene oxide group, which is significantly smaller than the value observed in our experiments (270 $\mu$S/cm per mol of ethylene oxide group in Brij). This value is substantial higher than the one found for ethylene oxide in surfactants like Tween 60, which shows and increase of 130 $\mu$S/cm per mol of ethylene oxide group. Therefore, the observed increase in conductivity in pure Brij may be attributed to the superpolarization effect and the presence of impurities in the samples.
The other curves in   \autoref{Fig5}a are the electrical conductivity subtracted by the contribution of ionization of water by Brij. 

For all samples, it is observed a non-trivial behavior of successive drops and rises of the electrical conductivity as more Brij is added to the solution. This effect is magnified for more concentrated solutions of SDS. From the previous sections we were able to elaborate some hypotheses to explain each of the sequential behavior, as follows:

\vspace{0.1cm}

\noindent	 \textbf{(i) \hspace{0.051cm} Decrease in electrical conductivity at low [Brij]:} At low [Brij] (from 0 to $\sim$ 10 mM, depending on [SDS]) a reduction in the conductivity is observed. The [Brij] at the most diluted solution is larger than its CMC ($\textrm{CMC}_{\textrm{Brij}} = 1.5 \times 10^{-2}$ mM) so it is expected that micelles of this surfactant are present from the beginning. It is reasonable to state that the addition of further Brij more micelles are formed and more SDS in solution are incorporated to these micelles followed by adsorption of part of the counterions. The consequence of both phenomena is the decrease in the electrical conductivity. The same phenomenon was observed in previous section with the decrease of conductometric CMC with increasing of [Brij] (\autoref{Fig2}e and \autoref{Fig4}b).
One may argue that the decrease in conductivity is solely due only to the micellization of Brij and SDS. However, the conductivity data from the samples with [SDS] = 18 mM without subtracting the signal of pure Brij shows a drop from 770 $\mu$S/cm to 690 $\mu$S/cm as the [Brij] increases from 0 to 15 mM (data not shown here). This decrease is significantly greater than the increase of conductivity observed for pure Brij over the concentration range of 0 to 70 mM. Therefore, the hypothesis of an alternative explanation than the micellization is consistent.

\vspace{0.1cm}	

\noindent	\textbf{(ii) \hspace{0.19cm} Increased in electrical conductivity:} With further addition of more Brij the counterions adsorbed in the micelles surface are released due to the weakening of the electrostatic force between micelle and counterions caused by the dilution of SDS molecules in the micellar phase. This behavior was discussed and observed in  \autoref{Fig4}c where the degree of ionization increases with [Brij] and in  \autoref{Fig2}c where it increases with the diminution of $\alpha$.

\vspace{0.1cm}	
\noindent	 \textbf{(iii) Decrease in electrical conductivity at high [Brij]:} It was assumed that the decrease in the electrical conductivity and so the charges' mobility at higher [Brij] is due to the increase of the viscosity of solution.

It is important to note that the addition of Brij triggers three phenomena: (i) the incorporation of SDS into the micelles; (ii) the dilution of SDS within the micelles and the release of counterions; and (iii) the increase in solution viscosity. These phenomena occur at all concentrations of Brij, although the dominance of each phenomenon varies with the concentration range.

To evaluate the extension of the phenomenon (\textbf{i}) the [Brij] at the minimal electrical conductivity, $\textrm{[Brij]}_{\kappa_{\mbox{{\scriptsize min}}}}$, was plotted against the [SDS]. From this point the phenomenon (\textbf{ii}) should start to be dominant. In the same way the [Brij] that maximizes the electrical conductivity, $\textrm{[Brij]}_{\kappa_{\mbox{{\scriptsize max}}}}$, marks the point from which the phenomenon (\textbf{iii}) starts to be dominant. The results are presented in  \autoref{Fig5}b.

It can be seen that $\textrm{[Brij]}_{\kappa_{\mbox{{\scriptsize min}}}}$ increase with increasing [SDS] until [SDS] $\sim$ 8 mM (value close to the CMC of SDS). This can be explained by the presence of more SDS in the solution, more counterions, increasing the range of [Brij] in which occurs the association of SDS and adsorption of counterions (\textbf{i}). For [SDS] $>$ 8 mM (larger than the CMC) the extension of the first phenomenon is reduced since part of SDS molecules are already at the micellar form and less free SDS and counterions are available to be incorporated in the mixed micelles. This assumption is corroborated by reference \cite{quina1995growth}  that shows a reduction around 1 mM of free SDS concentration when the [SDS] increases from 8 mM to 20 mM.

On the other hand, $\textrm{[Brij]}_{\kappa_{\mbox{{\scriptsize max}}}}$ increases monotonically with [SDS], but apparently changing its slope around 8 mM. In this case, the increase SDS molecules in the solution also increases a fraction of SDS in the micellar phase. Therefore, the release of counterions by dilution of SDS in mixed micelles is also expected to increase monotonically. The observed slope change maybe indicates the formation of mixed micelles rich in SDS as observed by the conductometric CMC in  \autoref{Fig4}b. 

In order to evaluate the magnitude of each phenomenon, we calculated the variation in electrical conductivity at each phenomenon as follows: $\Delta \kappa \textbf{i} = \kappa_{\mbox{{\scriptsize 0}}} - \kappa_{\mbox{{\scriptsize min}}}$; $\Delta \kappa \textbf{ii} = \kappa_{\mbox{{\scriptsize max}}} - \kappa_{\mbox{{\scriptsize min}}}$ and $\Delta \kappa \textbf{iii} =  \kappa_{\mbox{{\scriptsize max}}} - \kappa_{\mbox{{\scriptsize 0}}}$, where $\kappa_{\mbox{{\scriptsize 0}}}$ is the conductivity at [SDS] = 0 mM and the subscript max and min stands for the local maximum and local minimum. 

It can be seen in  \autoref{Fig5}c that $\Delta \kappa\textbf{i}$ has a similar behavior than $\textrm{[Brij]}_{\kappa_{\mbox{{\scriptsize max}}}}$ but a more pronounced change also at [SDS] $\sim$ 8 mM. The same interpretation could be given here: the more SDS in solution, the more counterions and more pronounced is the drop of $\kappa$. The reduction of this parameter can be explained by the existence of SDS micelles prior to the addition of [Brij].

The $\Delta \kappa\textbf{ii}$ increases linearly with the [SDS], which is explained by the increases of SDS density on the micellar phase and increases on the amount of released counterions by SDS dilutions at micelles with the addition of Brij.
This effect is analogous to the reduction in the degree of ionization of the mixed micelles with the increase of [SDS] (\autoref{Fig2}c) explained by the increase in the density of SDS in the micelles.

To understand the behavior of $\Delta \kappa\textbf{iii}$ imagine a large amount of Brij added to the system. In this scenario the mixed micelles should have a very few electrical charges in their surface since [SDS] $\ll$ [Brij]. So no counterions should be attached to the micelles and the conductivity should be close to the initial conductivity $\kappa_{\mbox{{\scriptsize 0}}}$, assuming that there will be no increase in the viscosity of the solution with the addition of Brij. In this case, $\kappa_{\mbox{{\scriptsize max}}}=\kappa_{\mbox{{\scriptsize 0}}}$, $\Delta \kappa\textbf{iii} = 0$ and thereby $\Delta \kappa\textbf{ii} = \Delta \kappa\textbf{i}$. The fact that $\kappa_{\mbox{{\scriptsize max}}}$ is smaller than $\kappa_{\mbox{{\scriptsize 0}}}$, and so $\Delta \kappa\textbf{iii} < 0$  implies that even if all counterions are free there is a decrease in their mobility when Brij is added. We associate this with the effect of partially impeding this mobility due to the increased viscosity of the solution. It is worth noting that for [SDS] $> \textrm{CMC}_{\textrm{SDS}}$, o $\Delta \kappa\textbf{iii}$ is less negative. This is because the initial conductivity $\kappa_{\mbox{{\scriptsize 0}}}$ is reduced due to the adsorption of part of the counterions in the SDS micelles. This behavior can be seen in \autoref{Fig5}c.

\begin{figure}[H]
\centering
\includegraphics[width=\linewidth]{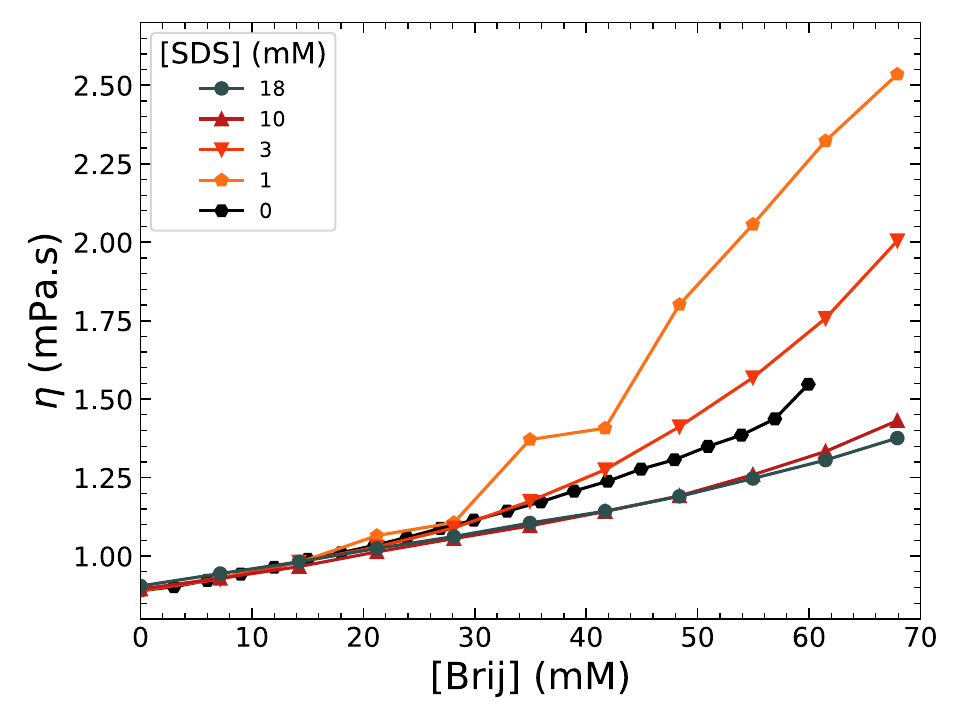}
\caption{Dynamic viscosity of solutions with [SDS] kept constant and varying [Brij]. The temperature of the system was kept constant ($25.0 \pm 0.1$) $^{\circ}$C.
}
\label{Fig6}
\end{figure}

To test the hypothesis of the effect of viscosity on the conductivity the dynamic viscosity ($\eta$) of solution changing [Brij] keeping [SDS] constant (see  \autoref{Fig6}) were made (\autoref{calcvisc}). The density of the solutions varied from 994.8 ${\textrm{kg}}/{\textrm{m}^3}$ at [Brij] = 0 mM to 996.8 ${\textrm{kg}}/{\textrm{m}^3}$ at [Brij] = 68 mM. The densities for intermediate concentrations were obtained by linear interpolation of these limit values.

The viscosity increases as more Brij is added and it is independent of [SDS] at [Brij] $<$ 30 mM. At [Brij] $>$ 28 and 35  mM it can be seen a strong increase in the viscosity for dilute solutions of SDS at 1 mM and 3 mM. In this same range small values of $\Delta \kappa\textbf{ii}$ is observed (\autoref{Fig5}c) suggesting that the viscosity effect is stronger in those concentrations. No strong increase of viscosity was observed at concentrations higher than the CMC of pure SDS ([SDS] = 10 and 18 mM). This corroborates the hypothesis of the effect of viscosity, since the release of counterions of the step (ii) should have a weaker effect in the conductivity for small [SDS] when compared to the viscosity effect.

However, we have no explanation for the sudden increase of viscosity for small [SDS] after $\sim$ 28 mM of Brij. One possible explanation is that more complex micellar structures are formed, such as wormlike micelles \cite{yue2009phase}. This hypothesis was discarded since the concentration of both surfactants to form this kind of micelles is at least 2 orders of magnitude higher than the ones we used here \cite{yue2009phase}.

Another phenomenon that can explain the transition on viscosity is the electroviscous effect, which is an increase in the viscosity caused by the attraction and alignment of water molecules by molecules with high enough charges \cite{uchiyama1990viscosities}. The electroviscous effect could explain the viscosity transition observed on [SDS] = 1 and 3 mM since the higher point of counter-ion liberation from micelles ($\textrm{[Brij]}_{\kappa_{\mbox{{\scriptsize max}}}}$) is at the same order of magnitude of the increase in $\eta$. The same transition should be observed for [SDS] = 10 and 18 mM at [Brij] above 70 mM from which the [Brij] should be sufficient to release the counterions adsorbed in micelles, making the electroviscous effect significant.

\section{Conclusion}
\label{Conclusion}

In this work, the thermodynamic behavior of aqueous solutions containing the surfactants SDS and Brij L4 was systematically studied by electrical conductivity, tensiometry and viscosimetry. Even though ionic and nonionic micellar systems have been extensively studied \cite{rony2023influences,li2016mixtures,kumar2024binary,zhang2005interaction,mazen2008mixing,
ABBOT2021116352,sahu2015anionic, quina1995growth,uchiyama1990viscosities,yue2009phase}, this study provides new insights through three distinct approaches: varying [SDS] at constant [Brij]; varying [Brij] at constant [SDS] and varying the total surfactant concentration keeping the ratio of the two surfactants constant. 
It was determined that tensiometry gives the true CMC (the onset of micelle formation) while conductometry gives the transition from which the micelles has a sufficient amount of SDS to enhance the micellar surface charge. It was found that interaction between the surfactants is synergistic with $\beta = -3.2$ and the molar concentration of SDS in the micelles became significant when SDS corresponds to 90 \% of the surfactants in solution or higher. 
In all cases, the presence of Brij decreases CMC and increases the degree of ionization. 
In the experiments of electrical conductivity keeping the [SDS] constant it was observed a sequential effect of decreasing and increasing of electrical conductivity as more Brij was added to the solution. This result was not reported in previous works and was associated to the following phenomena: i) the incorporation of free SDS molecules into the mixed micelles; ii) the dilution of SDS molecules within the micelles and the release of adsorbed counterions; iii) an increase of the solution viscosity. Electroviscosity effects were observed as the viscosity of the mixture was strongly increased with more Brij for small SDS amounts.

\section*{Credit authorship contribution statement}

\textbf{Juliano F. Teixeira:} Methodology, data curation, investigation and formal analysis of all experiments and data. Writing – original draft, review \& editing. Visualization.
\textbf{Juliana S. Quintão:} Formal analysis of experiments and data. 
\textbf{Kairon M. Oliveira:} Formal analysis of experiments and data. 
\textbf{Alvaro V. N. C. Teixeira:} Supervision, data curation, investigation, formal analysis. Review \& editing. Project administration. Funding acquisition.

\section*{Acknowledgements}
This research was funded by FAPEMIG (Fundação de Amparo à Pesquisa do Estado de Minas Gerais, project APQ-02733-18) and CNPq (Conselho Nacional de Desenvolvimento Científico e Tecnologico). J. F. Teixeira expresses his gratitude to CAPES (Coordenação de Aperfeiçoamento de Pessoal de Nível Superior) for the study grant. The author are thankful to M. R. Tótola (head of Laboratório de Biotecnologia e Biodiversidade para o Meio Ambiente) and E. Basílio de Oliveira (LOP/LEMA) from Universidade Federal de Viçosa.


\bibliographystyle{unsrt}
\bibliography{Bibliography}

\end{multicols}

\newpage
\begin{center}
\section*{\Large Supplementary Material}
\end{center}
\label{Supplementary Material}

\section*{Constant ratio of surfactants}

\begin{figure}[H]
\centering
\includegraphics[width=1\linewidth]{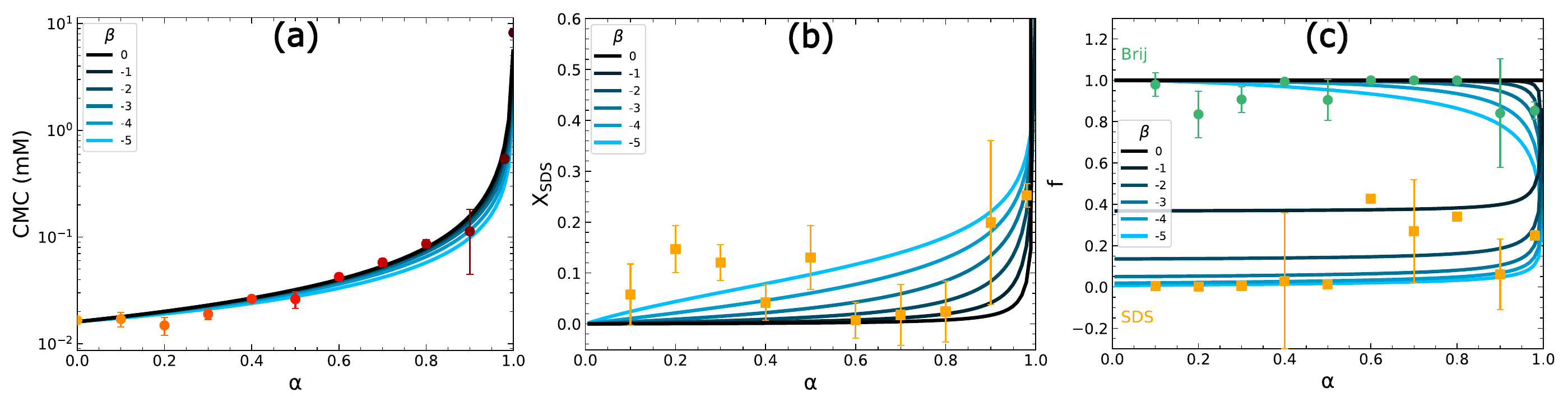}
\caption{\textbf{(a)} Theorical behavior for differents $\beta$ values. \textbf{(b)} SDS and Brij activity coefficients for differents $\beta$ values. 
}
\label{Fig7}
\end{figure}

\autoref{Fig7}a shows the theorical behavior of CMC varying $\alpha$ for differents $\beta$. The difference between ideal mixing and non-ideal mixing is indeed small and seems negligible. However when we  analyse  but also the SDS molar fraction inside the micelles and the activity coefficients (\autoref{Fig7}b and \autoref{Fig7}c) the difference is more clear. We considere the theorical behavior for differents $\beta$ and $\beta = -3.2$ is the value that better represents the interaction between the surfactants.

The \autoref{Tab1} correspond to interfacial tension data.

\begin{table}[H]
\centering
\caption{CMC of mixed surfactant solution, molar fraction of SDS on solution ($\alpha$), molar fraction of SDS at the mixed micelle ($X_{\textrm{SDS}}$), interaction parameter ($\beta$) and activity coefficients of SDS and Brij ($f_{\textrm{SDS}}$, $f_{\textrm{Brij}}$).}

   \begin{tabular}{>{\centering\arraybackslash}p{0.3cm}  >{\centering\arraybackslash}p{2.63cm}  >{\centering\arraybackslash}p{1.8cm}  >{\centering\arraybackslash}p{1.8cm}  >{\centering\arraybackslash}p{2.63cm}  >{\centering\arraybackslash}p{2.21cm} }
  \hline
\textbf{$\alpha$} &	CMC (mM)	&  \textbf{$X_{\textrm{SDS}}$} & \textbf{$\beta$} & \textbf{$f_{\textrm{SDS}}$} & \textbf{$f_{\textrm{Brij}}$}	\\  \hline

0		&	$0.0166 \pm 0.0009$	&	-				 &	-				&	-					&		-			\\ [2pt]

0.1		&	$0.017 \pm 0.003$	&	$0.06 \pm 0.06$ &	$-6 \pm 2$	&	$0.004 \pm 0.003$	&	$0.98 \pm 0.06$	\\ [2pt]

0.2		&	$0.015 \pm 0.003$	&	$0.15 \pm 0.05$	&	$-8 \pm 1$	&	$0.0024 \pm 0.0009$	&	$0.8 \pm 0.1$	\\ [2pt]

0.3		&	$0.019 \pm 0.002$	&	$0.12 \pm 0.04$	&	$-6.7 \pm 0.9$	&	$0.006 \pm 0.002$	&	$0.91 \pm 0.06$	\\ 

0.4		&	$0.026 \pm 0.001$	&	$0.04 \pm 0.04$	&	$-4 \pm 2$		&	$0.0 \pm 0.3$		&	$0.99 \pm 0.01$	\\ 

0.5		&	$0.026 \pm 0.005$	&	$0.13 \pm 0.06$	&	$-6 \pm 2$		&	$0.01 \pm 0.01$		&	$0.9 \pm 0.1$	\\ 

0.6		&	$0.042 \pm 0.003$	&	$0.01 \pm 0.04$	&	$-1 \pm 3$		&	$0 \pm 5$			&	$1.00 \pm 0.01$	\\ 

0.7		&	$0.058 \pm 0.005$	&	$0.02 \pm 0.06$	&	$-1.4 \pm 0.8$	&	$0.3 \pm 0.2$		&	$1.000 \pm 0.001$	\\ 

0.8		&	$0.086 \pm 0.008$	&	$0.02 \pm 0.06$	&	$-1 \pm 2$		&	$0 \pm 2$			&	$0.999 \pm 0.002$	\\ 

0.9		&	$0.11 \pm 0.07$		&	$0.2 \pm 0.2$	&	$-4 \pm 3$		&	$0.1 \pm 0.2$		&	$0.8 \pm 0.3$	\\ 

0.98	&	$0.54 \pm 0.05$		&	$0.25 \pm 0.02$	&	$-2.5 \pm 0.3$	&	$0.25 \pm 0.02$		&	$0.85 \pm 0.04$	\\ 

1		&	$8.42 \pm 0.01$	&	-				 &	-				&	-					&		-			\\ 
\hline
\multicolumn{6}{c}{$\beta$ 	average: $-4 \pm 1$	}				\\ 
\hline
\end{tabular}
\label{Tab1}
\end{table}

\section*{Thermodynamic characterization}

\begin{figure}[h]
\centering
\includegraphics[width=\linewidth]{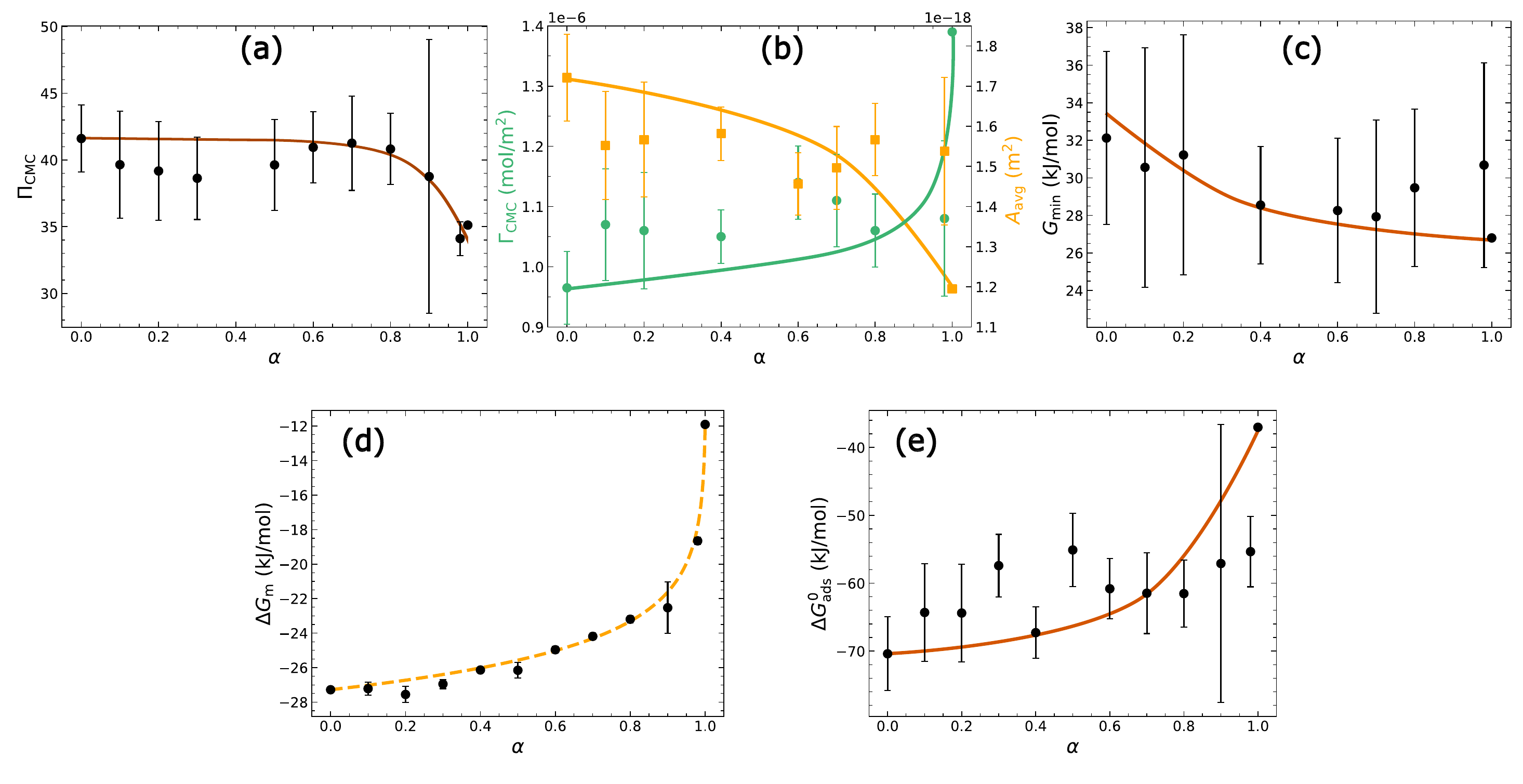}
\caption{Thermodynamic parameters vs $\alpha$: \textbf{(a)} surface pressure; \textbf{(b)} surface excess concentration ($\Gamma_{\textrm{CMC}}$) and average area occupied by a single surfactant molecule at the interface ($A_{\textrm{avg}}$); \textbf{(c)} variation of free energy at the air/water interface; \textbf{(d)} variation of the free energy change of micellization. The dashed line represents the theoretical behavior of $\Delta G_{m}$ for ideal mixing given by Regular Solution Theory. \textbf{(e)} Variation of the free energy change of adsorption. All continuous lines are guides to the eye.
}
\label{Fig8}
\end{figure}

Superficial activity of surfactants can be evaluated by calculation of surface pressure in air/water interface at the CMC values ($\Pi_{\textrm{CMC}}$), as follows by \autoref{eqS1}:

\begin{equation}
\Pi_{\textrm{CMC}} = \gamma(0)-\gamma(\textrm{CMC})
\label{eqS1}
\end{equation}

\noindent where  $\gamma(0)$ is the interfacial tension of the pure solvent and $\gamma(\textrm{CMC})$ is interfacial tension at CMC. $\gamma(\textrm{CMC})$ values were determined by  \autoref{eqTI} and the results of $\Pi_{\textrm{CMC}}$ are shown in  \autoref{Fig8}a.

\autoref{Fig7}a show that surface pressure for  $\alpha = 0$ is bigger than for $\alpha = 1$ denoting that nonionic surfactant Brij reduces the interfacial tension  more than anionic surfactant SDS.

Despite the scattering of the points, the results qualitatively follow the result observed in a similar system, as show in reference \cite{sahu2015anionic}.

In addition to surface pressure, we can determine the concentration of surfactant adsorbed at the air/water interface at the CMC. This concentration is expressed by surface excess concentration $\Gamma_{\textrm{CMC}}$ using  \autoref{atsup}.

\begin{equation}
\Gamma_{\textrm{CMC}} = \frac{1}{2.303nRT} \left(\frac{d\gamma}{d\textrm{log}([S])}\right)_{T} 
\label{atsup}
\end{equation}

\noindent where $R = 8.314$ J/(mol.K) is the universal gas constant, $T$ is the temperature, $\left(d\gamma/d\textrm{log}([S])\right)_{T}$ correspond to the slope near the CMC on a graphic of interfacial tension ($\gamma$) versus log concentration and $n$ is a constant related with the adsorption of charges at the interface.

For solution containing only nonionic surfactant $n=1$, if we have only anionic surfactant $n=2$. For surfactants mixtures with both types of surfactants, we have $n=3$ \cite{sahu2015anionic}.

The derivative $\left(d\gamma/d\textrm{log}([S])\right)_{T}$ was calculated using  \autoref{eqTI}, obtained by fit the experimental data showed in  \autoref{Fig7}b, which is given by the average of the slopes before and after the CMC. 

From the surface excess concentration values given in mol/$\textrm{m}^2$, we can determine the average area in $\textrm{\AA}^{2} $ occupied by a single molecule ($A_{\textrm{avg}}$) at the interface:

\begin{equation}
A_{\textrm{avg}} = \frac{10^{20}}{N_{A} \Gamma_{\textrm{CMC}}}
\label{Amedi}
\end{equation}

\noindent where  $N_{A} = 6.022 \times 10^{23}$ $\textrm{mol}^{-1}$ is the Avogadro's number. The results of $A_{\textrm{avg}}$ for different $\alpha$ value are shown in  \autoref{Fig7}b.

Surface excess concentration $\Gamma_{\textrm{CMC}}$ and average area $A_{\textrm{avg}}$ for different $\alpha$ values are shown in \autoref{Fig7}b.
$\Gamma_{\textrm{CMC}}$ values are positive for all $\alpha$ indicating that surfactants are favorably adsorbed at the interface.
We also see that $\Gamma_{\textrm{CMC}}$ increase and $A_{\textrm{avg}}$ decrease with increase $\alpha$ value, showing that the anionic surfactant has a surface excess bigger than nonionic. This means that anionic surfactant decreases more the interfacial tension of solution. This also explain the SDS average area is smaller than Brij due to the affinity of SDS on interface it is such that the activity is greater and the area smaller, even with electrostatic repulsion.

The fact of hydrophobic chain of SDS is the same of the Brij, the results show us the hydrophilic chain effect on $\Gamma_{\textrm{CMC}}$ and $A_{\textrm{avg}}$. The effect it is related with steric repulsion  super molecular polarization in Brij hydrophilic chain  \cite{duan2011super,duan2012super}, making the  $\Gamma_{\textrm{CMC}}$ of  Brij smaller and $A_{\textrm{avg}}$ bigger than SDS. It is important to emphasize that the surface excess is related to the facility with which surfactant decreases the interfacial tension with increasing surfactant concentration. On the other hand, the surface pressure expressed up to the amount of interfacial tension that reduced to the CMC point. Thus, the results show that the surface pressure of SDS is lower than that of Brij surfactant and that SDS more easily reduces the interfacial tension of the solution.

To evaluate the synergism of the mixture, we determined the free energy at the air/water interface ($G_{\mbox{{\scriptsize min}}}$), given by  \autoref{eqgmin}. The free energy at the interface correspond to the energy needed to transfer 1 mole of surfactant ($\Gamma_{\textrm{CMC}}$) from the bulk of solution to the interface.

\begin{equation}
G_{\mbox{{\scriptsize min}}} = A_{\textrm{avg}} \gamma(\textrm{CMC}) N_A .
\label{eqgmin}
\end{equation}

\noindent The values of $G_{\mbox{{\scriptsize min}}}$ for different $\alpha$ values are presented in  \autoref{Fig7}c.

As $G_{\mbox{{\scriptsize min}}}$ decreases with the increasing of $\alpha$, is more thermodynamically favorable to transfer the SDS surfactant from the bulk to the interface. This means that in a mixture system the presence of SDS in the interface is preferable and greater than Brij.

To evaluate the micellization in a system with mixture of surfactant, we determined the thermodynamic parameter   $\Delta G_{m}$ related with variation on free energy during the micellization, as follows:

\begin{equation}
\Delta G_{m} = RT\ln(\textrm{CMC}).
\end{equation}

In this calculation, the adsorption of counterions by micelles was neglected because the anionic surfactant concentration was small. In \autoref{Fig7}d we have $\Delta G_{m}$ for different $\alpha$ values.  

All values $\Delta G_{m}$ are negatives (\autoref{Fig7}d) meaning that as expected the micelles' formation  are thermodynamically favorable. 
The values of $\Delta G_{m}$ increases with increasing $\alpha$ values, indicating that micelle formation is less favorable as $\alpha$ increases. This behave is due to electrostatic repulsion between the molecules of SDS surfactant.

Also, we calculated the free energy change of adsorption ($\Delta G_{\textrm{ads}}$) given by  \autoref{eqGads}.

\begin{equation}
\Delta G_{\textrm{ads}} = \Delta G_{m} -  \Pi_{\textrm{CMC}}/\Gamma_{\textrm{CMC}}
\label{eqGads}
\end{equation}

\noindent Results of $\Delta G_{\textrm{ads}}$ are shown in \autoref{Fig7}e.
We can see that $\Delta G_{\textrm{ads}}$ became less negative with increasing $\alpha$. This behave is related to the adsorption of Brij in mixed micelles, decreasing the electrostatic repulsion between SDS molecules.

\end{document}